# Individualized Algorithmic Advice as a Strategic Signal on Competitive Markets☆

Tobias R. Rebholz [a,b], Maxwell Uphoff [a,c], Christian H. R. Bernges [a], and Florian Scholten [a]

[a] Department of Psychology, University of Tübingen
[b] Fuqua School of Business, Duke University
[c] Department of Economics, University of Minnesota Twin Cities

## Abstract

As algorithms increasingly mediate competitive decision-making, their influence extends beyond individual outcomes to shaping strategic market dynamics. In our experiment, we examined how algorithmic advice affects human behavior in a classic economic game with a unique, non-collusive, and analytically traceable equilibrium. Participants ($N$ = 129) played a Cournot quantity competition with equilibrium-aligned or strategically biased algorithmic recommendations. While individualized equilibrium advice supported stable convergence, collusively downward-biased advice led to sustained underproduction and supracompetitive profits—hallmarks of tacit collusion. Participants' quantities converged faster and more consistently toward individualized than collective equilibrium advice, potentially due to an objective quality advantage or greater perceived ownership of the former. These findings demonstrate that algorithmic advice can function as a strategic signal, shaping coordination even without explicit communication. The results echo real-world concerns about algorithmic collusion and underscore the need for careful design and oversight of algorithmic decision-support systems in competitive environments.

---

☆ The data, materials, surveys, and analysis scripts for the main study (Cournot quantity competition) and the pilot study (Bertrand price competition) are publicly available at the Open Science Framework (https://osf.io/sw9g2). The full experimental instructions can be found in the Online Appendix. *Conflicts:* The authors have no known conflicts of interest to disclose. *Funding:* The authors' contributions to this research were supported by the following institutions. T.R.R.: German Research Foundation (Walter Benjamin grant no. 553942000), Universitätsbund Tübingen e.V. (research grant no. 5095); F.S.: German Research Foundation (GRK 2277 "Statistical Modeling in Psychology," project no. 310365261), German Academic Exchange Service (RISE Germany grant no. CS-BI-4992). *CRediT:* The authors made the following contributions. T.R.R.: Conceptualization (equal), Data Curation (lead), Formal Analysis (lead), Funding acquisition (lead), Investigation (lead), Methodology (lead), Project Administration (lead), Resources (supporting), Software (supporting), Supervision (equal), Validation (lead), Visualization (lead), Writing - Original Draft, Writing - Review & Editing (equal); M.U. and C.B.: Conceptualization (equal), Data Curation (supporting), Formal Analysis (supporting), Funding acquisition (supporting), Investigation (supporting), Methodology (supporting), Resources (lead), Software (lead), Validation (supporting), Visualization (supporting), Writing - Review & Editing (equal); F.S.: Conceptualization (supporting), Formal Analysis (supporting), Funding acquisition (supporting), Investigation (supporting), Methodology (supporting), Project Administration (supporting), Resources (supporting), Software (supporting), Supervision (equal), Writing - Review & Editing (equal). *Corresponding author:* Tobias R. Rebholz, 100 Fuqua Drive, Durham, NC 27708, USA. Email: tobias.rebholz@duke.edu

## 1. Introduction

Algorithmically generated advice—such as output from generative AI (GenAI), including large language model-based chatbots—is becoming an increasingly pervasive force in individual and organizational decision-making. From dating apps and media recommendations to financial forecasting and price optimization, algorithms are shaping how humans make judgments and act upon information. Correspondingly, the field of algorithmic advice taking—the study of how humans interpret, trust, and act on algorithmic suggestions—has gained growing attention (see Burton et al., 2020; Jussupow et al., 2020; Mahmud et al., 2022, for reviews).

The primary focus of research on algorithmic advice taking has been on non-strategic, individual decision contexts, where the outcome for the advisee and final decision-maker is independent of the behavior of other advisees. In many of these settings, individuals have exhibited *algorithm aversion*—a tendency to discount or distrust algorithmic recommendations relative to those of other humans (Mahmud et al., 2022; Meehl, 1954; Prahl & van Swol, 2017), especially when exposed to visible errors of both types of advisors (Dietvorst et al., 2015). In the absence of performance information, other work has found *algorithm appreciation*, with people preferring algorithmic over human advice across a broad range of individual decision-making contexts (Logg et al., 2019). Recent evidence in economic tasks also raises concerns about overreliance on algorithmic advice (Tse et al., 2024). Factors such as task type (e.g., objective vs. subjective), expertise of advisors and advisees, algorithmic transparency, and general familiarity with AI systems have been identified as critical moderators of this behavioral polarity (Castelo et al., 2019; Mahmud et al., 2022).

However, many real-world applications of algorithmic decision-making unfold in strategic environments, where outcomes are contingent on the interdependent decisions of multiple agents—such as in rental pricing, auction bidding, or production settings. In these domains, algorithms not only assist individual decisions but simultaneously alter the behavioral ecology (i.e., the structure of interactions and adaptive behaviors shaped by the environment; Hutchinson & Gigerenzer, 2005) of entire markets by influencing how agents anticipate and respond to one another. Specifically, when market participants expect others to rely on algorithmic advice to a certain extent, they may adjust their own adherence accordingly, leading to systemic changes in patterns of interaction and overall market dynamics. This raises important and underexplored questions about how humans integrate algorithmic advice in strategic settings. While individual contexts often focus on the trustworthiness and usefulness of advice for the decision-maker's own outcome, the function of algorithmic advice in multi-agent settings shifts: The potential of an algorithm's strategic signal to facilitate competition or coordination ultimately depends not only on decision-makers' own dispositions (i.e., appreciation vs. aversion), but also on their beliefs about competitors' inclination to interpret and use algorithmic advice. Our main results suggest that humans treat algorithmic advice not merely as recommendations, but as strategic signals—factoring in what others might infer and how they may react to individual or shared advice.

### *1.1. Algorithmic Collusion*

Concerns about algorithmic collusion—where algorithmic advice leads independent actors to align their decisions in ways that reduce competition—have intensified in recent years. A prominent real-world case involves RealPage Inc.'s rental pricing algorithm which was adopted by many large landlords in the United States. By providing rent recommendations based on aggregated competitor data, RealPage allegedly enabled property managers to raise rents in lockstep, thus essentially suppressing price competition without direct coordination (Kaye, 2024; U.S. Department of Justice, 2024). It has been argued that their algorithm facilitated a form of hub-and-spoke collusion: A behavioral ecology in which a central entity (the "hub;" here: RealPage, acting as a data-analytics provider) facilitates coordination among otherwise independent actors (the "spokes;" here: the landlords) through indirect means (e.g., algorithmic price recommendations; Harrington, 2026). Whether the resulting alignment of actors' strategies is intentional or not is irrelevant to the occurrence of algorithmic collusion.

The case of RealPage exemplifies that, under certain conditions, algorithmic advice has the potential to function beyond its role as a mere instrument to facilitate decision-making processes. It can become a strategic signal, shaping expectations about other individuals' behavior, with corresponding consequences for the adaptation of one's own behavior. Consequently, economists and antitrust experts have warned that the widespread deployment of AI-powered decision-support systems may facilitate non-transparent forms of collusion, particularly in oligopolistic markets (Assad et al., 2024; Calvano et al., 2020). This is corroborated by experimental evidence that humans indeed tacitly collude with algorithmic systems (Zhou et al., 2018), and by simulations demonstrating how algorithms can learn to collude with one another (Calvano et al., 2020; see also March, 2021, for a review). However, empirical research directly examining whether and how humans coordinate their behavior with each other in response to algorithmic advice—particularly when such advice is provided to multiple participants in the same strategic setting—remains limited.

Recent experiments study algorithmic price recommendations (Hunold & Werner, 2025), including in human-only and hybrid human-AI markets (Schauer & Schnurr, 2023), and the delegation of pricing decisions to an algorithm (Normann et al., 2025). However, algorithmic collusion is not unique to pricing but can similarly occur for algorithmic advice on strategic business decisions for which the collusion mechanisms are less obvious than elevating prices in lockstep. Specifically, in contrast to price competition, market outcomes in quantity competition depend on total output, and supracompetitive outcomes emerge indirectly through sustained underproduction rather than directly through elevated prices. As described in more detail in the following section, our contribution thus focuses on participants' autonomous production decisions. Moreover, instead of investigating algorithmic agency, we focus on algorithmic guidance: Participants receive recommendations, but final quantity choices remain human. Therefore, algorithmic collusion would be the consequence of human behavior (advice taking) rather than algorithmic actions. Put differently, we are interested in the influence of decision-support systems on market outcomes in explicit contrast to full automation of strategic decision processes.

### *1.2. The Present Study*

In this paper, we present evidence from a preregistered experiment that aims to fill this gap in the literature by examining how algorithmic advice influences human decision-making in strategic games. Participants interacted in a well-established economic game derived from industrial organization theory, where the strategic dynamics are analytically traceable and equilibrium solutions are unique, non-collusive (i.e., competitive), and well understood. We manipulated whether participants received algorithmic advice, as well as whether the advice was individualized (i.e., private information) or collective (i.e., shared across participants). We also manipulated whether the advice was aligned with the Nash equilibrium or strategically biased toward monopoly outcomes (i.e., fostering tacit collusion). Our aim is to compare how different forms of algorithmic advice influence the collective dynamics of coordination. Whereas advice presumably serves as a strategic signal in all conditions, only collusive advice is strategically biased in the sense of *systematically favoring its recipients* over third parties—especially consumers, who must pay supracompetitive prices, but also other suppliers in markets without access to such algorithmic advice.

Our study contributes to the literature on augmented judgment and decision-making in several respects. First, it extends traditionally individual-focused research on (algorithmic) advice taking into strategic, multi-agent settings. Second, we compare the effects of individualized versus collective advice. This contrast is conceptually relevant because collective advice can act as a shared focal signal, whereas individualized advice is private and therefore cannot serve as the same common coordinating device. It is also practically relevant because it resembles the transition from more traditional (deterministic) to contemporary (stochastic) decision-support systems. Specifically, collective advice is intended to mimic the functioning of deterministic decision-support systems (presumably the RealPage algorithm), which thus generate output that is shared across users. In contrast, contemporary GenAI systems produce context-sensitive responses that vary with the prompts provided by individual users, resulting in inherently individualized output from algorithmic advisors such as large language model-based chatbots (Costello et al., 2024; Leung & Urminsky, 2025; see also Bubeck et al., 2023; Vaswani et al., 2017). Third, we test whether biased algorithmic advice can induce strategically meaningful behavioral shifts, including signs of tacit collusion. Collective advice biased toward monopoly outcomes can facilitate price collusion among suppliers (Hunold & Werner, 2025). In contrast, receiving slightly different (i.e., individualized) recommendations makes it more difficult for suppliers to coordinate their strategies. In other words, individualized algorithms remove coordination as a motive for following their advice, helping us to single out the effect of people’s dispositions in relying on decision-support systems.

In our experiment as described in more detail below, participants played a Cournot competition assuming the roles of firms setting production quantities (Holt, 1985; Huck et al., 1999, 2002, 2004). To test the technical implementation of algorithmic advice in strategic games, we also conducted a pilot study using Bertrand price competition (Dufwenberg et al., 2007; Dufwenberg & Gneezy, 2000, 2002; see the Online Appendix for details). The strategic dynamics of Cournot competition are generally less prone to collusion than for Bertrand competition (Melkonyan et al., 2018; Suetens & Potters, 2007; Zhou et al., 2018). In repeated Cournot markets with more than two firms and outcome feedback, naïve participants without decision support usually begin by producing quantities that oscillate around the Nash equilibrium (Rassenti et al., 2000). Learning the strategic dynamics of the game thus implies their quantities asymptotically approaching the optimal quantity, what we refer to as convergence.

The advice algorithm generated recommendations aligned with the best-response dynamics of the game. That is, it was based on the game-theoretic best response to what one's competitors did in the previous period (see Melkonyan et al., 2018, for details). Therefore, we expect that participants in conditions receiving such advice will converge to the stable equilibrium faster than naïve participants in the control condition without advice (Hypothesis 1). By design, participants' strategies and the advice converge in tandem to the equilibrium over the course of the experiment. As a result, participants' quantities should increasingly match their received advice (Hypothesis 2). In other words, even though the algorithms' recommendations lag behind by one period, participants should learn that incorporating them is beneficial for their individual profits. However, as long as not all participants have reached equilibrium, best responses to other participants' behavior are not necessarily shared. The best response-based generation mechanism presumably implies a quality advantage for the strategic signal provided by individualized advice: It conveys more tailored and thus individually valuable information than collective advice, which is generated by averaging the best responses of all firms in the market. Therefore, we expect that the rate at which participants converge to individualized advice over the course of the experiment will be greater than the convergence rate in the collective advice condition (Hypothesis 3).

Beyond private versus shared information, in the individualized advice condition we additionally distinguished between equilibrium-aligned and collusively biased algorithms. Providing individualized recommendations that align with the best-response dynamics of the game should promote convergence to the equilibrium as described above. In contrast, the collusive advice-generation mechanism was designed to induce tacit collusion by altering the individual best response in a way to provide monopoly-aligned (i.e., downward-biased) quantity recommendations. Therefore, we expect that participants in the collusive advice condition will systematically underproduce, that is, converge to smaller average production quantities than participants in other conditions (Hypothesis 4). As the magnitude of bias was set equal to the distance between the equilibrium and monopoly quantities, following such advice would yield supracompetitive profits (i.e., profits that are higher than what could be sustained in a fully competitive market). Accordingly, we also expect participants in the collusive advice condition to achieve higher profits than participants in other conditions (Hypothesis 5). Note that any production quantity below the equilibrium level is considered collusive (due to resulting in supracompetitive market prices; Suetens & Potters, 2007). In other words, stable convergence to the collusive quantity is not necessary to demonstrate algorithmic collusion. Even if participants only partially undercut the equilibrium in this condition, this would harm market efficiency: The surplus of hypothetical consumers is reduced in proportion to the extent that our participants acting as suppliers on the market follow the collusively biased advice.

## 2. Method

In each period, participants in fixed groups of three produce quantities between 0 and 100. The market price is a decreasing function of the total production quantity in the market, resulting in a unique Nash equilibrium at 24.75 quantity units per participant and a collusive monopoly production quantity of 16.50 units. This game structure usually features initial overproduction (e.g., Holt, 1985), making strategic advice particularly valuable to reduce overall production. Quantity recommendations evolved in response to participants' previous behavior and were either individualized or collective. In addition, we distinguished between equilibrium-aligned and collusively biased individualized algorithms. This allowed us to test whether participants would follow advice that systematically deviates from the game-theoretic equilibrium in a direction that benefits all group members, but harms market efficiency by reducing the consumer surplus. Taken together, this study enables us to explore not only whether participants heed algorithmic advice, but also whether they coordinate around it when it is strategically biased in their favor.

This experiment was preregistered (https://osf.io/qa4yc) and conducted online using the LIONESS Lab platform (Giamattei et al., 2020). Deviations from the preregistration are documented in the Online Appendix, and the data, materials, surveys (incl. full experimental instructions), and analysis scripts can be found in the Open Science Framework repository (https://osf.io/sw9g2). We report how we determined our sample size, all data exclusions (if any), all manipulations, and all measures.

### *2.1. Participants*

In line with our preregistered resource constraints, we aimed to recruit the maximum number of participants feasible within the available project budget. A total of 185 participants were recruited via Prolific and took part online between February 17–18, 2025. All participants were prescreened to be fluent English speakers from the UK. They were compensated with a base payment of £2 for a study lasting approximately 20 minutes, with the opportunity to earn an additional bonus via a lottery system. Each participant's total profits accumulated over the entire game determined the number of lottery tickets for a raffle for one of four £15 bonus payments. Of those who entered the study, 132 reached the matching stage, where they were grouped with two other participants. One group of three participants in the collusive advice condition was excluded due to incorrect advice display caused by an unknown technical problem, yielding a final sample size of $N = 129$ (gender: 66 female, 63 male; age: $M = 35.40$, $SD = 12.40$) for analysis. This corresponds to 43 independent groups and 3,225 individual-period observations.

### *2.2. Design*

The experiment employed a mixed design with 25 repeated measures (quantity production periods) and one between-subjects factor (type of algorithmic advice: individualized equilibrium vs. collective equilibrium vs. individualized collusive vs. none). Due to resource constraints, a full crossing of aggregation (individualized vs. collective) and alignment (equilibrium vs. collusive) was inadmissible, so the collective collusive advice condition was omitted from the design matrix. In order to be able to compare the effects of algorithmic advice to a naïve baseline, no advice was provided to participants in the control condition. In the advice conditions, participants received period-specific quantity recommendations generated by an algorithm simulating strategic learning. That is, algorithmic recommendations were aligned with game-theoretic best responses to observed competitor behavior from the previous period. In the experimental instructions, the algorithm was described at a high level as training on the decisions of all players. In our implementation, this corresponds to updating recommendations each round via a best-response computation based on prior-round quantities, plus mean-zero noise (see below for more details). In the collusive condition, however, advice was systematically biased downward (by 8.25 units, which is exactly the difference between the Nash and collusive quantities) to nudge behavior toward tacit collusion. Participants were only informed whether their advice was individualized or shared. They were not informed about treatment differences in whether the advice was equilibrium-aligned versus strategically downward-biased.

### *2.3. Materials and Procedure*

Participants played 25 periods of a homogeneous Cournot competition game (Huck et al., 1999, 2002, 2004). In each period, three participants acted as symmetric firms in a closed market, producing quantities ranging from 0 to 100 in 0.01 increments. The market price was determined endogenously as the maximum quantity minus the total quantity (or 0, whichever was greater). Participants' profits were calculated as the market price minus a constant cost of 1 multiplied by their individual quantities. This game has a unique Nash equilibrium at a quantity of 24.75 units per firm, yielding a total output of 74.25 units and a market price of 25.75. Moreover, a collusive monopoly outcome exists at 16.50 units per firm, yielding a total output of 49.50 units and a market price of 50.50. Thus, any amount of underproduction relative to equilibrium indicates collusive tendencies proportional to the degree of supracompetitiveness of the resulting market prices (Suetens & Potters, 2007).

Before any data were collected, participants were informed about the general purpose of the research, any foreseeable risks or benefits of the tasks to be completed, and that participation was voluntary and anonymous. However, they were not informed about all features of the advice-generation mechanism (e.g., whether recommendations were equilibrium-aligned or strategically biased). Participation was voluntary and anonymous and could not have caused any long-term effects. All participants provided written consent to participate in the study and to publish their anonymized data. Then, they read detailed instructions and completed comprehension checks requiring manual calculation of potential market outcomes, designed to reinforce the strategic logic of the game as described in the instructions. Upon passing the checks, participants were randomly assigned into groups of three and played the full 25 periods without dropout. At the end of each period, participants received feedback on their group's quantities, the resulting market price, and their individual profits. The profits of the two competitors were not displayed directly but could be inferred.

In the advice conditions, participants were told they would receive algorithmically generated recommendations based on prior market behavior. Specifically, they were told that "the algorithm will pick an initial value in the first round and train on the decisions of all players in subsequent rounds of play in this game." In the first period, advice was actually drawn from a truncated normal distribution ($M = 55$, $SD = 10$), serving as a high anchor to induce slow convergence to equilibrium. From the second period onward, advice was dynamically computed as the best response to the competitors' behavior in the previous period, with added white noise ($M = 0$, $SD = 2$) to introduce variability and avoid fully deterministic recommendations across periods (cf. how "creativity" is simulated in GenAI; Bellemare-Pepin et al., 2024; Davis et al., 2024). This stochastic component avoids giving participants a perfectly predictable mechanical rule, which helps reduce purely calculative responding. The experimental instructions explicitly paralleled the stochasticity in the advice generation mechanism with the functioning of GenAI: "As with other algorithms (e.g., ChatGPT), the advice generated by this algorithm is not always the same. Every round, it will pick a number from its distribution."

In the individualized equilibrium advice condition, each participant received a unique quantity recommendation based on their individual best-response function. In the collective equilibrium advice condition, all participants received the same quantity recommendation calculated as the average of quantities resulting from the three individual best-response functions. In the individualized collusive advice condition, each participant's unique best-response function was altered by applying a systematic downward bias of 8.25 units, the result of which was provided as an algorithmic quantity recommendation. Thus, participants in this condition received advice nudging them toward tacit collusion. Importantly, however, as the algorithm still followed best-response logic, it also recommended individual defection in underproductive regimes, mimicking the instability of collusion. Participants were informed about the individualized versus shared nature of their advice. Specifically, the instructions at the beginning of the experiment read: "Note that in each round, the quantity recommended by the algorithm is chosen [individually / collectively] for all three players. So each player has access to [separate, individual / the same] advice." This information was repeated on each of the 25 screens where participants were asked to set their production quantities.

### *2.4. Measures and Data Analysis*

The primary behavioral measure was the quantity produced by each participant in each period. Additional derived measures included individual profits and the absolute deviations of participants' quantities from equilibrium (24.75), monopoly (16.50), and the recommended quantity (varying across periods). After the game, participants completed a short questionnaire assessing their familiarity with GenAI, trust in the algorithm (for those in advice conditions), and provided open-ended feedback.

For the main analyses of Hypotheses 1–3, we estimated mixed-effects exponential decay models of the form

$$y_{igt} = \left(a_0 + a_g + \sum_k a_k \mathbb{I}_k\right) \exp(-(b_0 + \sum_k b_k \mathbb{I}_k)t) + (c_0 + \sum_k c_k \mathbb{I}_k) + \varepsilon_{igt} \quad (1)$$

with group $g$-wise random intercepts $a_g \sim N(0, \sigma^2_{a,g})$ to account for within-group interdependences (i.e., repeated interactions across periods $t = 0, \dots, 24$). We modeled separate parameters per condition $k \in \{IEA, ICA, CEA\}$ for the intercept $a$, decay rate $b$, and asymptote $c$, where $\mathbb{I}_k$ denote dummy indicator variables (with $\mathbb{I}_{IEA} = 1$ for individualized equilibrium advice, $\mathbb{I}_{ICA} = 1$ for individualized collusive advice, $\mathbb{I}_{CEA} = 1$ for collective equilibrium advice, and 0 otherwise). For the main analyses of Hypotheses 4–5, we estimated linear mixed-effects regression models of the form

$$y_{igt} = \beta_0 + \alpha_g + \beta_t t + \sum_k \beta_k \mathbb{I}_k + \sum_k \beta_k (t \times \mathbb{I}_k) + \varepsilon_{igt} \quad (2)$$

with group $g$-wise random intercepts $\alpha_g \sim N(0, \sigma^2_{\alpha,g})$ and interactive fixed effects of periods $t = 0, \dots, 24$ and condition $k \in \{NAC, IEA, CEA\}$. Again, $\mathbb{I}_k$ are dummy indicator variables (with $\mathbb{I}_{NAC} = 1$ for the no advice control, $\mathbb{I}_{IEA} = 1$ for individualized equilibrium advice, $\mathbb{I}_{CEA} = 1$ for collective equilibrium advice, and 0 otherwise).

## 3. Results

### *3.1. Descriptive Analyses*

We first examined how participants' production quantities evolved over the 25 periods of the Cournot competition across the four conditions (see Figure 1, Panel A). In the initial periods, quantities in the advice conditions were higher than in the control condition, indicating early adherence to upward-biased algorithmic advice (see also Figure 1, Panel C). Over the course of the experiment, however, quantities in the advice conditions decreased, reflecting convergence toward the Nash equilibrium of 24.75 units (see Figure 1, Panel A). Unexpectedly, participants in the control condition started near the equilibrium but gradually diverged in the direction of the competitive Walrasian quantity of 33, where the price drops to 1 and firms produce without profit (see also Huck et al., 1999). Therefore, participants in the equilibrium advice conditions quickly overtook the control group, remaining consistently closer to the equilibrium in later periods for individualized advice. At least partially consistent with Hypothesis 1, the advice-generation mechanism following best-response dynamics appeared to successfully guide these participants toward the Nash equilibrium. As a consequence of their tandem convergence, and in line with Hypothesis 2, the deviation between participants' quantities and the algorithmic advice mostly *decreased* from period to period (see Figure 1, Panel D). Furthermore, descriptively consistent with Hypothesis 3, the rate of convergence appeared to be greater for individualized than for collective advice.

**Figure 1:**
Time Series of Average Quantities (A), Profits (B), Received Advice (C), and Deviation from Advice (D)

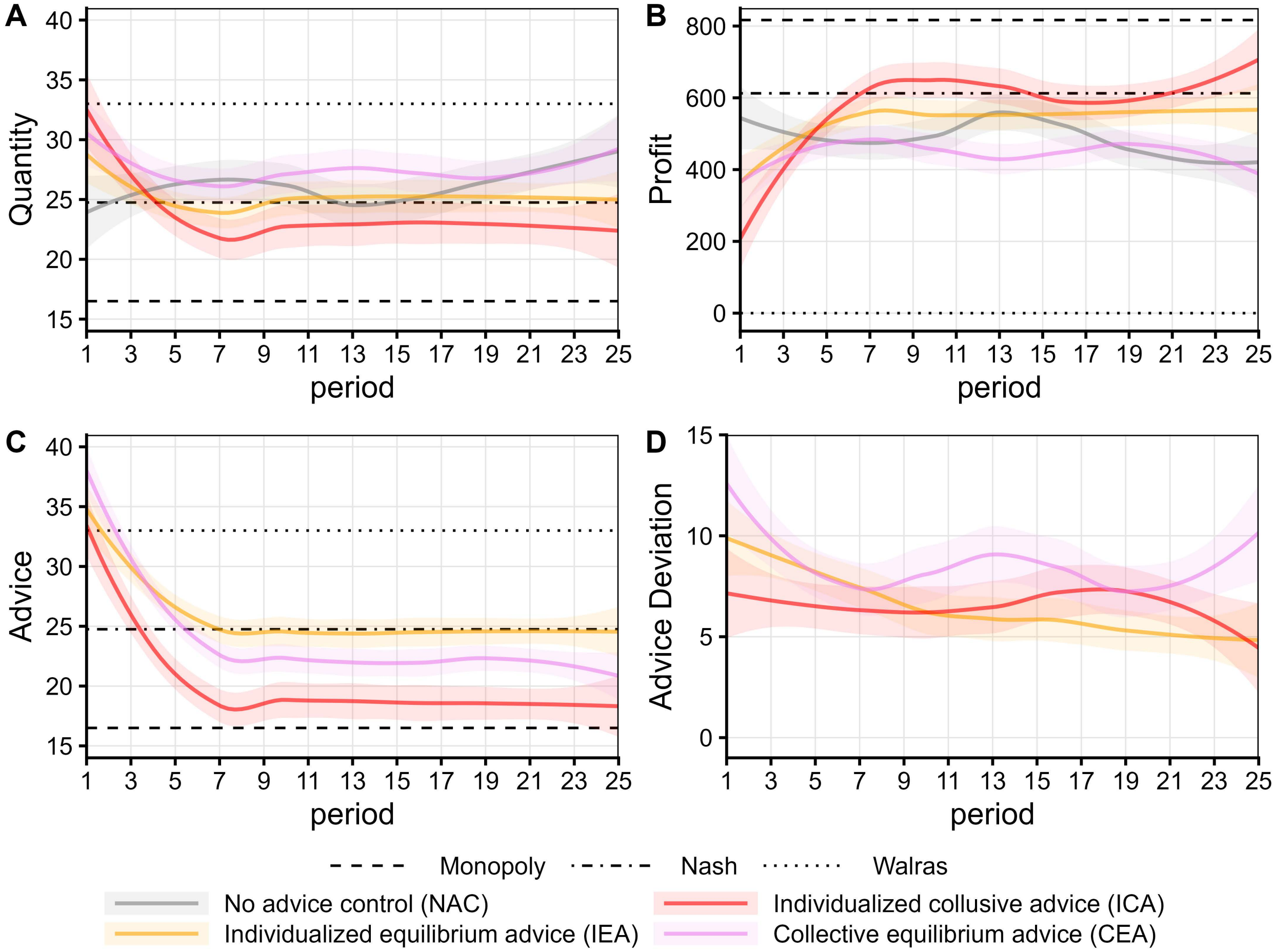

*Note.* The transparent areas around the solid local regression fits per condition indicate the 95% CI. The dashed lines show the monopoly quantity and profit, the dotted and dashed lines show the quantity and profit at the Nash equilibrium, and the dotted lines show the competitive Walrasian quantity and profit.

The individualized equilibrium advice condition exhibited the most stable and consistent convergence to the Nash equilibrium, whereas the collective equilibrium advice condition showed greater variability and a somewhat unexpected tendency toward slight overproduction despite average advice to underproduce (see Figure 1, Panel A vs. C). Most importantly, and in line with Hypothesis 4, participants in the individualized collusive advice condition consistently produced quantities below the equilibrium benchmark in later periods. Although they did not fully converge to the collusive outcome of 16.50 units, there was sustained underproduction in this condition. Consistent with Hypothesis 5, these participants benefited from higher profits relative to participants in other conditions (see Figure 1, Panel B). In general, the profit trajectories closely paralleled the production quantity patterns. Participants in the individualized collusive advice condition achieved the highest profits by producing below the equilibrium level. Participants in the individualized equilibrium advice condition also experienced gradual increases in profits, reflecting improved strategic alignment with the equilibrium quantity, whereas overproduction had negative effects on profits in the collective equilibrium advice and control conditions, in line with the observed divergence in the direction of competitive Walrasian quantities in later periods.

To contextualize these observations, supplementary analyses considered two counterfactual benchmarks. Specifically, the outcome trajectories for the hypothetical scenario of all participants perfectly following collective equilibrium advice reveal that this would indeed have resulted in underproduction and thus higher profits than perfectly following individualized equilibrium advice (see Online Appendix, Figure S2, Panel B). In contrast, the hypothetical scenario of all participants perfectly implementing the best response to their competitors' behavior reveals that joint defection from individualized collusive advice would have resulted in the lowest average profits across all conditions (see Online Appendix, Figure S3, Panel B), highlighting the instability of tacit algorithmic collusion.

For the questionnaire at the end of the survey, Cronbach's $\alpha$ was adequate (mean inter-item $r$ = .70, $\alpha$ = .88, 95% CI [.83, .92], $n$ = 96) for the three items measuring trust in the specific algorithm used in the three advice conditions ("I think the algorithm used in the task would act in a way that is advantageous for me;" "I trust the algorithm used in the task;" and "The algorithm used is well equipped for this game;" all measured on 5-point scales from 1 = "Strongly disagree" to 5 = "Strongly agree"). Therefore, we created an algorithm trust index by averaging the responses to all three questions. This index revealed moderate levels of trust in the algorithm ($M$ = 3.20, $SD$ = 1.02), which did not differ significantly across conditions ($M_{IEA}$ = 3.29, $SD_{IEA}$ = 1.19; $M_{ICA}$ = 3.23, $SD_{ICA}$ = 1.07; $M_{CEA}$ = 3.07, $SD_{CEA}$ = 0.80; all pairwise $|t(93)| \leq 0.88$, Tukey-adjusted $p$s > .050). In this questionnaire, participants also reported descriptively higher perceived individualization for individualized than for collective algorithms ($M_{IEA+ICA}$ = 3.27, $SD_{IEA+ICA}$ = 1.15; $M_{CEA}$ = 3.09, $SD_{CEA}$ = 1.04; $t(94) = 0.75$, $p$ = .458) in response to the item "The algorithm responded well to my individual behavior" (measured on the same 5-point scale from 1 = "Strongly disagree" to 5 = "Strongly agree").

### *3.2. Confirmatory and Exploratory Inferential Analyses*

To statistically test differences in production quantity trajectories, we conducted mixed-effects exponential decay modeling of the deviation of participants' quantities from the equilibrium, estimating separate parameters per condition (see Table 1, Column 1). The initial amplitudes in the advice conditions were significantly higher than in the control condition, consistent with the initial upward bias programmed into the advice-generation mechanism. Although participants in the control condition started closer to equilibrium, those in the advice conditions ended up closer due to the observed divergence in the absence of advice. There was a significant negative decay rate in the control group, which provided statistical evidence for the unexpected increase in production quantities over the course of the experiment. Most importantly, however, this model revealed significantly higher and positive decay rates for all advice conditions. In other words, there was evidence for Hypothesis 1 that algorithmic advice fostered convergence toward equilibrium. Note that Wald $\chi^2$ tests of linear restrictions on the fixed effects suggested that there were no significant differences in decay rates between individualized equilibrium and collusive advice, $b$ = -0.24, $\chi^2(1)$ = 1.63, $p$ = .201, individualized and collective equilibrium advice, $b$ = 0.08, $\chi^2(1)$ = 0.29, $p$ = .593, or individualized collusive and collective equilibrium advice, $b$ = 0.32, $\chi^2(1)$ = 3.58, $p$ = .058.

**Table 1:**

Mixed-Effects Exponential Decay Models of Absolute Deviations

| Parameter | Absolute deviation from equilibrium | Absolute deviation from advice |
|---|---|---|
| $a_0$ | 0.07 | 8.39*** |
| | (1.88) | (1.26) |
| $a_{IEA}$ | 12.80*** | |
| | (2.81) | |
| $a_{ICA}$ | 18.02*** | –1.87 |
| | (2.91) | (1.80) |
| $a_{CEA}$ | 11.61*** | –0.28 |
| | (2.79) | (1.75) |
| $b_0$ | –0.03** | 0.02*** |
| | (0.01) | (0.01) |
| $b_{IEA}$ | 0.65*** | |
| | (0.12) | |
| $b_{ICA}$ | 0.89*** | –0.02** |
| | (0.14) | (0.01) |
| $b_{CEA}$ | 0.57*** | –0.02*** |
| | (0.09) | (0.01) |
| $c_0$ | 9.65*** | |
| | (1.34) | |
| $c_{IEA}$ | –3.18* | |
| | (1.37) | |
| $c_{ICA}$ | –1.11 | |
| | (1.37) | |
| $c_{CEA}$ | –2.52 | |
| | (1.37) | |
| $\sigma_{a,g}$ | 5.44 | 3.66 |
| $N_g$ | 43 | 32 |
| $N_{total}$ | 3,225 | 2,400 |
| $R^2_{marginal}$ | 0.11 | 0.01 |
| $R^2_{conditional}$ | 0.18 | 0.14 |

*Note.* Standard errors are in brackets. The models correspond to Equation (1). In Column 1, the no advice control condition is the reference group (IEA: Individualized equilibrium advice; ICA: Individualized collusive advice; CEA: Collective equilibrium advice). In Column 2, $c = 0$ and the individualized equilibrium advice condition is the reference group. Two-sided *** $p < .001$, ** $p < .01$, * $p < .05$.

To more directly evaluate the extent of advice taking, we also conducted mixed-effects exponential decay modeling of the absolute deviation between participants' quantities and the algorithmic advice (see Table 1, Column 2). Convergence issues did not allow us to estimate free asymptote parameters $c$. Refraining from doing so is appropriate here because quantities and advice converge in tandem by design. Consistent with Hypothesis 2, the significant positive intercept of the decay parameter indicated that participants' quantities and the individualized equilibrium advice converged toward each other. However, the significant negative interaction terms for this parameter were similar in magnitude to the intercept. Thus, there was also evidence for Hypothesis 3, or a significantly greater rate of convergence toward individualized equilibrium advice than toward collective equilibrium advice. Indeed, Wald $\chi^2$ tests of linear restrictions on the fixed effects did not provide evidence for non-zero net decay rates for both individualized collusive advice, $b = -0.0003$, $\chi^2(1) = 0.0027$, $p = .959$, and collective equilibrium advice, $b = -0.0026$, $\chi^2(1) = 0.3246$, $p = .569$, indicating no meaningful convergence in these conditions.

To examine whether collusive behavior emerged in the individualized collusive advice condition as predicted, we used linear mixed-effects regression analyses predicting participants' quantities and profits as functions of period, condition, and their interaction (see Table 2, Columns 1 & 2). For quantities in the collusive advice condition, there was a significant negative effect of period, indicating decreasing quantities under systematically downward-biased advice, consistent with Hypothesis 4. To evaluate whether these trajectories translated into lower end-of-experiment quantities, we conducted Wald $\chi^2$ tests of linear restrictions on the fixed effects with period set to its maximum value. These tests indicated that participants in the individualized collusive advice condition finished the experiment with significantly lower production quantities than those in the no advice control condition, $\beta = -6.13$, $\chi^2(1) = 13.35$, $p < .001$, the individualized equilibrium advice condition, $\beta = -3.40$, $\chi^2(1) = 4.12$, $p = .042$, and the collective equilibrium advice condition, $\beta = -5.93$, $\chi^2(1) = 12.48$, $p < .001$. Indeed, the reduced quantities in the collusive advice condition were accompanied by a significant positive effect of period on profits, indicating a supracompetitive advantage consistent with Hypothesis 5. Corresponding Wald $\chi^2$ testing of linear restrictions at the final period further indicated that the individualized collusive advice condition finished the experiment with significantly higher profits than the no-advice control condition, $\beta = 265.23$, $\chi^2(1) = 17.41$, $p < .001$, and the collective equilibrium advice condition, $\beta = 252.19$, $\chi^2(1) = 15.74$, $p < .001$, but only descriptively higher profits than the individualized equilibrium advice condition, $\beta = 109.20$, $\chi^2(1) = 2.95$, $p = .086$. This overall pattern is consistent with strategic convergence to a collusive regime, albeit informally and without explicit coordination.

**Table 2:**
Linear Mixed-Effects Models of Strategies and Outcomes

| Parameter | Quantity | Profit |
|---|---|---|
| $\beta_0$ | 25.38*** (1.21) | 464.96*** (46.01) |
| $\beta_{NAC}$ | –0.52 (1.68) | 64.47 (63.57) |
| $\beta_{IEA}$ | –0.06 (1.68) | 24.92 (63.57) |
| $\beta_{CEA}$ | 1.79 (1.68) | –11.87 (63.57) |
| $\beta_t$ | –0.16** (0.06) | 9.81*** (1.62) |
| $\beta_{t \times NAC}$ | 0.28*** (0.08) | –13.74*** (2.24) |
| $\beta_{t \times IEA}$ | 0.14 (0.08) | –5.59* (2.24) |
| $\beta_{t \times CEA}$ | 0.17* (0.08) | –10.01*** (2.24) |
| $\sigma_{\alpha,g}$ | 2.78 | 126.62 |
| $N_g$ | 43 | 43 |
| $N_{total}$ | 3,225 | 3,225 |
| $R^2_{marginal}$ | 0.02 | 0.03 |
| $R^2_{conditional}$ | 0.07 | 0.17 |

*Note.* Standard errors are in brackets. The models correspond to Equation (2), where periods are denoted as $t = 0, \ldots, 24$ and the individualized collusive advice condition is the reference group (NAC: No advice control; IEA: Individualized equilibrium advice; CEA: Collective equilibrium advice). Two-sided *** $p < .001$, ** $p < .01$, * $p < .05$.

## 4. Discussion

We examined how algorithmic advice influences behavior in strategic environments where individuals' outcomes depend not only on their own decision but also on the strategies of others. The results of our experiment indicate that participants initially follow the algorithms' quantity recommendations—particularly when they are individualized—and adjust their behavior over time as they learn the structure of the game. Individualized equilibrium advice leads to more stable convergence to the game-theoretic equilibrium compared to collective advice or no advice. More importantly, this experiment demonstrates that algorithmic advice can also bias behavior in a coordinated direction, with significant implications for market outcomes. In the individualized collusive advice condition, participants consistently underproduce and achieve above-equilibrium profits—hallmarks of tacit collusion. The subtle and sustained bias of the algorithm toward lower quantities apparently shapes participants' behavior in a collectively profitable direction.

### *4.1. Practical Implications*

For markets with more than two firms, it has been shown that communication is essential for the emergence of collusion (see Schwalbe, 2018, for a review). However, our findings provide direct evidence that biased or opaque algorithms can act as catalysts for anticompetitive regimes even without explicit coordination or communication among human actors. Individualized algorithmic advice alone can apparently create conditions fostering de facto collusive outcomes. Taken together, our results highlight the complex and context-sensitive function of algorithmic advice in strategic environments, where its influence can either enhance efficiency or promote undesirable coordination, depending on how it is structured. Our study contributes to a growing literature on human-algorithm interaction by showing that the effects of algorithmic advice extend beyond individual accuracy or adherence. In fact, algorithmic decision-support systems can shape strategic dynamics and collective outcomes. The framing and structure of advice—individualized versus collective; equilibrium-aligned versus collusively biased—proved critical. Most importantly, participants' quantities converged faster and more consistently toward individualized, equilibrium-aligned quantity recommendations than toward collective ones.

The algorithmic advisor was explicitly described to participants as operating in a manner analogous to GenAI. This framing of recommendations sampled from a probability distribution is only a simplified proxy in the sense of sharing characteristics of contemporary GenAI systems, such as adaptivity and heterogeneity (Bellemare-Pepin et al., 2024; Davis et al., 2024). Participants' response to stochastic, GenAI-like advice as a meaningful strategic signal underscores the potential real-world impact of using such systems as decision-support tools in competitive contexts. For instance, this kind of technology is associated with serious risks, such as "hallucinations" (i.e., communicating false information with high confidence; Xu et al., 2024) and "metacognitive myopia" (i.e., lacking sensitivity to the history and validity of stimulus data; Scholten et al., 2025). Combined with the general "sycophancy" of many GenAI systems (i.e., their alignment with the primary goal of serving as helpful assistants; Sharma et al., 2024) and the highly idiosyncratic ways humans interact with them (e.g., confirmation-biased prompting; Leung & Urminsky, 2025; see also Bubeck et al., 2023), it is thus inevitable that different market participants will receive—at least slightly—different recommendations from GenAI. While this should theoretically make it more difficult for firms that increasingly rely on such systems for strategic planning to coordinate their strategies, our experiment nonetheless suggests that algorithmic decision-support systems that implement advice generation mechanisms akin to this kind of technology still pose a risk of algorithmic collusion.

Relatedly, the present advice-generation mechanism rests on a myopic forecasting assumption. Because the algorithm computes a best response to competitors' previous-period quantities, following its recommendation is strictly optimal only to the extent that rivals are expected to repeat, or approximately repeat, their most recent actions. If competitors instead anticipate future adjustments, deliberately experiment, or strategically deviate from prior strategies, the recommendation is no longer a fully optimal prescription but rather a psychologically plausible heuristic anchored in recent market behavior. We therefore interpret the present algorithm as capturing one tractable form of adaptive decision support in repeated competition, not as a globally optimal benchmark for all strategic environments. This boundary condition is practically important because it implies that the persuasive force of algorithmic advice may exceed its normative validity.

### *4.2. Limitations and Future Research*

We did not directly test any mechanisms that could explain faster and more consistent convergence toward individualized equilibrium advice than toward collective one. The most parsimonious explanation is that the word "individualized" signals higher quality compared to "shared" algorithmic advice, which should lead participants to weight the former more strongly than the latter (Patt et al., 2006; Yaniv & Kleinberger, 2000). Explicitly highlighting these properties of the decision-support systems in the instructions as done here may inadvertently have induced corresponding demand characteristics. To rule out this possibility, future research should employ more ecological settings with ambiguity about whether the algorithm acts on shared versus private information. Alternatively, individualized advice may be associated with a higher perceived ownership than collective advice. Consistent with this account, participants indeed reported descriptively higher perceived individualization for individualized than for collective algorithms. Accordingly, stronger weighting of advice from the former could be explained by the "mere ownership effect," the tendency to evaluate one's possessions more favorably or attribute higher psychological value to them (Beggan, 1992)—an interpretation that also warrants direct testing in future studies. More generally, future research should experimentally manipulate candidate mechanisms, including trust in the algorithm, perceived advice quality, perceived individualization or ownership, and expectations about competitors' use of the advice.

As implemented in the equilibrium advice conditions (except for the first period), some firms may indeed be motivated, willing, and able to elicit perfectly equilibrium-aligned advice from GenAI-based chatbots. However, their competitors may receive suboptimal or biased advice, for instance, due to algorithmic or human errors, biased output-generation mechanisms or misaligned prompting, insufficient context information, or the deliberate exploration or highlighting of alternative strategies. Moreover, this may be true only during some periods but not during others. These idiosyncrasies can create unpredictable dynamics affecting the interdependent outcomes of all market participants. In contrast, the constant strategic bias implemented in our algorithm is highly restrictive compared to truly adaptive advice generation mechanisms potentially affecting real-world markets. Moreover, because the collusive bias was only implemented for individualized advice, our design cannot determine if collective collusive advice would produce similar results. While this is highly plausible, given previous research demonstrating that collective advice can facilitate price collusion (Hunold & Werner, 2025), our findings should be interpreted as evidence that strategically biased individualized signals can facilitate underproduction and supracompetitive profits. They should not be interpreted as evidence that individualization is necessary for algorithmic collusion. Future studies should also include more ecological "mixed" groups, where participants in the same group receive different types of algorithmic advice, or are even allowed to freely prompt a GenAI for recommendations—presumably yielding maximally tailored, perfectly individualized algorithmic advice.

Several further limitations of our study merit discussion. First, our experimental design does not fully disentangle whether behavioral adjustments across periods reflect reactions to the algorithmic advice, the observed behavior of others, their interaction, or other factors (e.g., idiosyncratic learning styles or personality traits; Straub, 2009; Svendsen et al., 2013). Future studies could use adapted versions of our games—such as one-shot or sequentially forced-choice formats—to control for social learning, or include post-experimental surveys to probe motives and perceptions to gain clearer causal leverage. Second, we did not include punishment or enforcement mechanisms (e.g., explicit or implicit sanctions, institutional rules; cf. Hunold & Werner, 2025; Normann et al., 2025), or social norms (e.g., fairness, reputation) that might emerge in richer social environments (see Van Dijk et al., 2020, for an overview). For instance, introducing punishment opportunities in a public goods setting might help to address the issue that, at the aggregate level, many groups appeared to play an equilibrium strategy, whereas in reality a small number of participants exploited their competitors. Third, the current design could not distinguish whether advice indeed serves as a highly informative strategic signal or merely an uninformative anchor (cf. Hütter & Fiedler, 2019; Rader et al., 2015; Rebholz et al., 2026; Schultze et al., 2017). In fact, participants' self-reported trust in the algorithm measured at the end of the study was only moderate on average and did not show systematic differences across advice conditions. Accordingly, future studies should assess participants' beliefs and attributions regarding the value of advice more granularly across periods, focusing on the specific advice received (e.g., the perceived accuracy or informativeness of the recommended quantity) rather than on "the algorithm" more broadly.

Finally, our experiment involved only a single iteration of a multi-period game with a publicly known, deterministic termination point. Compared to real-world settings, our findings may thus be limited to competitions among a small number of startups with time-limited funding after a simultaneous launch. In contrast, research shows that experience affects participants' cooperative behavior (Dal Bó & Fréchette, 2011). This might also explain the unexpected divergence from the Nash equilibrium to Walrasian quantities over time in our control condition. Moreover, Cournot experiments vary in whether termination follows a publicly known finite, undisclosed, or stochastic rule, which may affect repeated-game incentives and observed convergence (Huck et al., 2004). Therefore, future research should use multiple iterations over uncertain time horizons to assess the stability of our findings among experienced players (i.e., established companies).

### *4.3. Conclusion*

The case of RealPage, a rental pricing algorithm allegedly used by large landlords to tacitly coordinate rent increases in U.S. housing markets (Kaye, 2024; U.S. Department of Justice, 2024), illustrates how biased algorithmic advice can effectively serve as a collusion-enabling signal in the real world. In our experimental Cournot market, participants sustainably aligned their behavior around a biased algorithmic recommendation, despite having no direct means of coordination (e.g., communication). As algorithmic decision-support tools become increasingly widespread in market settings, understanding how humans interpret and respond to these systems—not only in terms of individual decision-making but in shaping strategic outcomes—becomes ever more urgent. Our findings suggest that the framing and structure of algorithmic advice influence the cooperative dynamics these systems create in competitive environments. Most importantly, algorithmic decision-support systems that deliver highly individualized recommendations at scale—whether GenAI-based or not—may still pose a risk of inducing tacit collusion.